\documentclass[twocolumn,showpacs,preprintnumbers,amsmath,amssymb]{revtex4}

\usepackage{graphicx}% Include figure files
\usepackage{epsfig}% Include figure files
\usepackage{dcolumn}% Align table columns on decimal point
\usepackage{bm}% bold math

\begin{document}

%\preprint{As submitted to \prl}

\title{Large parallel and perpendicular electric fields \\on electron spatial scales
in the terrestrial bow shock
}% Force line breaks with \\

\author{S. D. Bale}
 \email{bale@ssl.berkeley.edu}
\author{F. S. Mozer}
\affiliation{%
Physics Department and Space Sciences Laboratory 
University of California, Berkeley, CA 94720}%

\date{March 8, 2007}% It is always \today, today,

\begin{abstract}
Large parallel ($\leq$ 100 mV/m) and perpendicular ($\leq$ 600 mV/m) electric fields were measured in the Earth's bow shock by the vector electric field experiment on the Polar satellite.  These are the first reported direct measurements of parallel electric fields in a collisionless shock.  These fields exist on spatial scales comparable to or less than the electron skin depth (a few kilometers) and correspond to  magnetic field-aligned potentials of tens of volts and perpendicular potentials up to a kilovolt.  
The perpendicular fields are amongst the largest ever measured in space, with energy densities
of $\epsilon_0 E^2/ n k_b T_e$ of  order 10\%.
The measured parallel electric field implies that the
electrons can be demagnetized, which may result in stochastic (rather than coherent) electron heating.
\end{abstract}

% PACS, 
\maketitle

In a collisionless plasma, shocks occur on scales much smaller than the mean free path for binary particle collisions.  In the case of
heliospheric and planetary bow shocks (Alfv\'en Mach number $M_A <$ 20), the macroscopic shock transition scale is approximately the gyroradius of ions trapped at the shock
front \cite{bale03}, which is order 1000 km, while the collisional mean free path is of order 1 AU (1.5 10$^{8}$ km).  Electric and magnetic
fields within the shock provide the required deceleration and entropy change \cite{goodrich84, scudder86}.  An important problem in shock physics is that of the energy budget across the shock.  A cold, super-Alfv\'enic upstream flow is thermalized as it crosses the layer; however the downstream partitioning of the available free energy between electrons, ions, and electromagnetic fields is not understood and is a key to interpreting observations of astrophysical shocks from
remote sensing measurements.  A poorly understood element in this energy partition problem is the cross-shock electric potential, how it arises and how it scales with
various shock parameters.
Collisionless shocks are also the source of the energetic particles ubiquitous in heliospheric and astrophysical contexts.  In
at least the case of the shock-surfing acceleration mechanism, the cross-shock electric field plays a fundamental role in controlling the
energization process \cite{shapiro03}.

The MHD generalized Ohm's law for an electric field in a plasma is  $\vec{E} + \vec{v}_e \times \vec{B} \approx  - \vec{\nabla}p_e/{ne} 
+ \eta \vec{j} + \frac{m_e}{ne^2}\frac{d\vec{j}}{dt}
$ (ignoring terms of order $m_e/m_i$) and indicates that any $E_\parallel$ is likely to arise from pressure or current gradients or anomalous resistivity.
In a deHoffman-Teller Lorentz frame, in which the incoming flow
and magnetic field are aligned, the convection electric field is zero.  Then electron acceleration and heating may be due to a parallel electric field $E_\parallel$.

The cross-shock direct current (DC) electric field is difficult to measure and has been reported rarely in the literature \cite{formisano82,
wygant87, scudder86, walker04} and parallel (magnetic field-aligned) electric fields have not been reported previously.
The electric field is known to be spiky, although previous DC measurements have often undersampled it.  AC-coupled spectral density and
waveform measurements measure broadband electrostatic turbulence \cite{rodriguez75, onsager89} that corresponds to Debye-scale 'electron holes' \cite{bale98}
and ion acoustic-like turbulence \cite{hull06}.

Here we report measurements from two crossings of the quasi-perpendicular terrestrial bow shock using data from the Polar satellite.  Polar has a 9.5 earth radius ($R_e$) apogee and encounters the bow shock in the sub-solar region only during periods of extreme solar wind pressure when the shock is compressed from its normal stand-off distance of $\sim$15 $R_e$.  Polar carries the only three-component DC electric field experiment that has been flown in the outer magnetosphere and bow shock.  It uses six spherical sensors, each of whose potentials are measured with respect to that of the spacecraft body.  The spheres are on the ends of booms such that, by pairs, they measure potential differences in the three orthogonal directions.  Spheres 1 and 2 are separated by 130 meters while spheres 3 and 4 are separated by 100 meters and are in the satellite spin plane.  Spheres 5 and 6 are aligned along the spin axis and are separated by 13.6 meters by rigid booms \cite{harvey95}.  Techniques for measuring parallel electric fields by such boom systems are described elsewhere\cite{mozer05}.  The electric field data described in this paper were collected in instrument burst modes at data rates of 1600 samples/second.  Vector magnetic fields are measured by the MFE fluxgate magnetometer instrument \cite{russell95} at a cadence of 8.3 vectors/sec. 

 \begin{figure}[h]
\centering
 \includegraphics[width=90mm,
	height=110mm, scale=1,clip=true, draft=false]{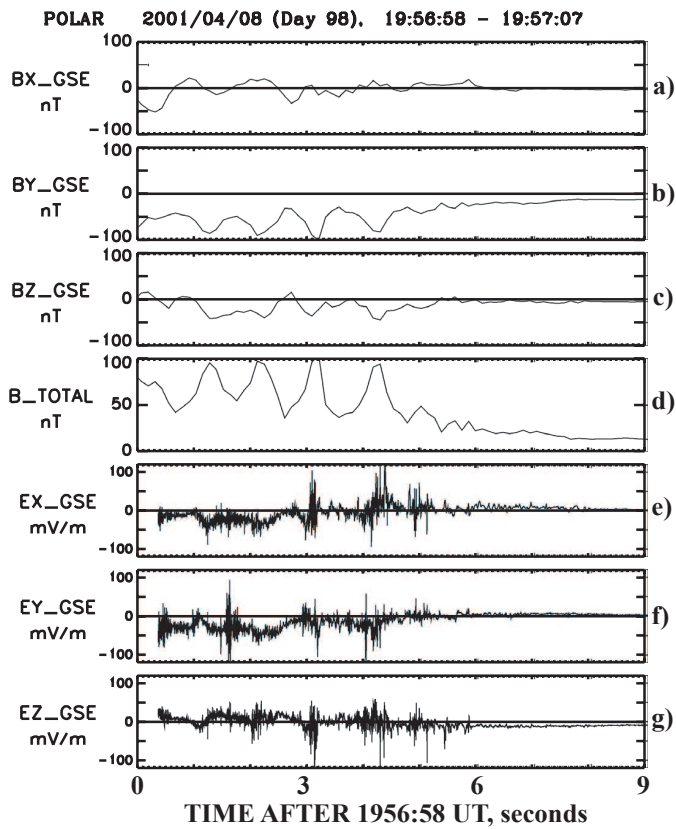}
\caption{Electric and magnetic field data in GSE coordinates during a Polar bow shock crossing on April 8, 2001.}
\end{figure}
     
          Fig. 1 shows nine seconds of magnetic and electric field data  on April 1, 2001 as the spacecraft crossed from the magnetosheath, through the bow shock, and into the solar wind near the subsolar point.  A shock normal is estimated by maximum variance of the electric field which agrees well with previous estimates
at this shock \cite{hull06}; shock parameters (estimated using ACE and Wind solar wind data convected to 1 AU) are shown in Table 1.
          The top three panels give the magnetic field components in GSE coordinates while panel d) gives the total magnetic field with several overshoots in the downstream region before decreasing to the solar wind value of about 15 nT.  The electric field components (in GSE) in the bottom three panels are characterized by $>$ 100 mV/m rapid oscillations that appear in the shock ramp.  Search coil magnetometer measurements at
these shocks (not shown) show broadband noise (probably whistler-mode), but {\em no} features that correspond to the electric fields reported here.   These electric fields
are therefore electrostatic and are either stationary structures within the shock, or moving very slowly with respect to it.

 \begin{figure}[h]
\centering
 \includegraphics[width=70mm,
	height=70mm, scale=1,clip=true, draft=false]{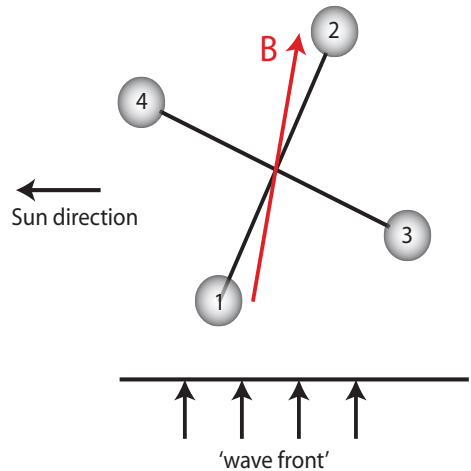}
\caption{Spin plane geometry of the electric field booms at 1856:58.55 UT on April 8, 2001.}
\end{figure}

From Fig. 2 it is seen that the electric field 'wave front' that passed over the sensors on April 8, 2001 at 18:56:58.55 UT traveled largely from sphere 1 to sphere 2 and nearly along the magnetic field direction.  Thus, signatures of the structure should be seen first in sphere 1 and then in sphere 2.  This expectation is borne out by the data of Fig. 3 where the sphere one signal in panel a) leads that from sphere 2 in panel b) by 1.8 data points (1.2 msec) according to their cross-correlation.  Otherwise the two signals are anti-correlated, signifying a large electric field in the 1-2 direction.  That the sphere 3 and sphere 4 signals are small indicates that the electric field was approximately parallel to the local magnetic field direction (see Fig. 2).  This expectation is borne out in the bottom three panels of Fig. 3, which give the electric field components in a magnetic field aligned coordinate system in which $\hat{z}$ is parallel to the local magnetic field, $\hat{x}$ is perpendicular to both the local magnetic field and to the vector from the spacecraft to the center of the earth, and $\hat{y}$ completes a right hand coordinate system.  From the time delay between the signals on spheres 1 and 2, the speed of the structure over in the spacecraft frame was 50 km/sec.  Thus, the total parallel field structure had a thickness of $\sim$2.5 km and the individual intensity peaks were $\sim$0.25 km thick.  The individual peak thicknesses were about equal to the gyroradius of a 25 eV electron and about 25\% of the electron skin depth, $c/\omega_{pe}$, where $\omega_{pe}$ is the electron plasma frequency.  The thickness of the entire 50 msec structure was about 5\% of the proton gyroradius or a few electron skin depths.  The net parallel potential across the structure was about 15 volts.  
The upstream plasma parameters are given in Table 1. 
     
 \begin{figure}[h]
\centering
 \includegraphics[width=90mm,
	height=110mm, scale=1,clip=true, draft=false]{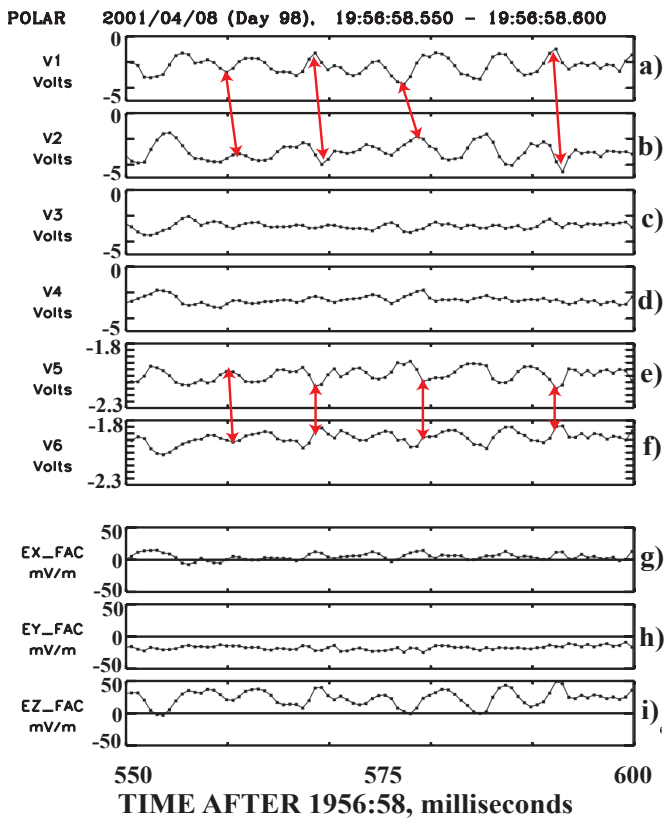}
\caption{Fifty milliseconds of field data in the bow shock ramp.  The top 6 panels are the single-ended probe
voltages and the bottom 3 panels are the electric field in a field-aligned system ($\hat{B} = \hat{z}$).}\label{fig:currents}
\end{figure}

          A second 50 msec example from the same shock crossing is given in Fig. 4, which has the same format as Fig. 3.  Because the time lag in the signals between opposite pairs of spheres in panels a) through f) was not measurably different from zero, the structure speed across the spacecraft was $>$50 km/sec.  For purposes of estimating measured parameters, the speed is assumed to be 50-100 km/sec
 \cite{bale03}.  In this example, perpendicular electric fields as large as 150 mV/m (panel (h)) and parallel fields as large as 100 mV/m (panel (i)) were measured, in the spacecraft frame.  Lorentz transformation to the shock frame (50-100 km/s) can give transformation fields as big as $|\vec{v} \times \vec{B}| \sim$ 10 mV/m to $E_\perp$; of course, the parallel electric field $E_\parallel = \vec{E} \cdot \hat{B}$ is a Lorentz invariant.  The 12 msec duration of these big field pulses corresponds to a thickness of 0.6-1.2 km, which is 0.5-1 times the electron skin depth and 0.3-0.5 times the radius of gyration of 10-50 eV electrons.  The total parallel electric potential across the structure was 30-60 volts.  
  
  \begin{figure}[h]
\centering
 \includegraphics[width=90mm,
	height=110mm, scale=1,clip=true, draft=false]{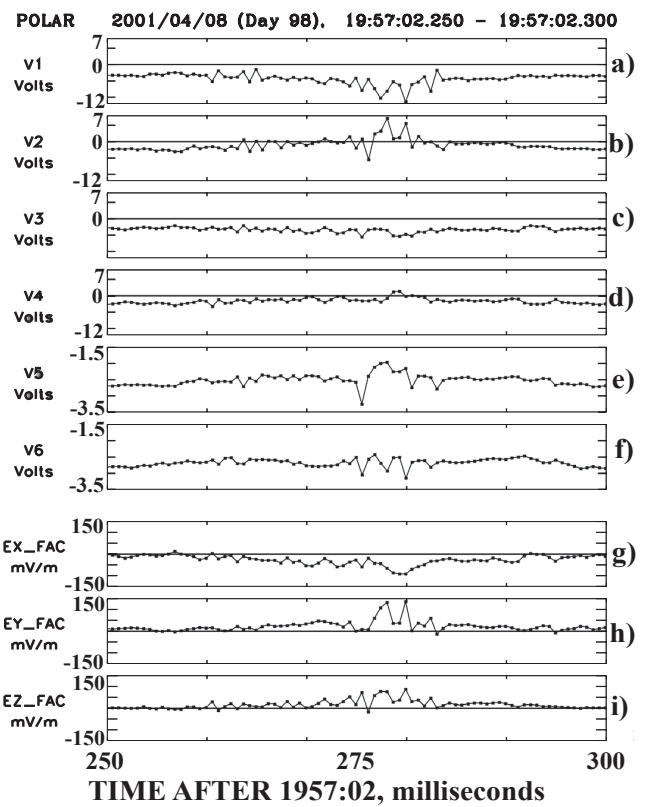}
\caption{Fifty milliseconds of field data at another interval during the April 8, 2001 bow shock in similar format
to Figure 3.}\label{fig:currents}
\end{figure}

       Another sub-solar quasi-perpendicular bow shock crossing occurred on April 11, 2001 and is shown in the one minute of data presented in Fig. 5.  Again, the spacecraft passed from the magnetosheath where the magnetic field intensity was $\sim$100 nT (panel d)) to the solar wind, where the field strength was about 25 nT.  And again, the electric field showed spiky structures (in the ramp but not in the solar wind) that exceeded the 200 mV/m dynamic range of the plot in panel e).
 
\begin{figure}[h]
\centering
 \includegraphics[width=90mm,
	height=110mm, scale=1,clip=true, draft=false]{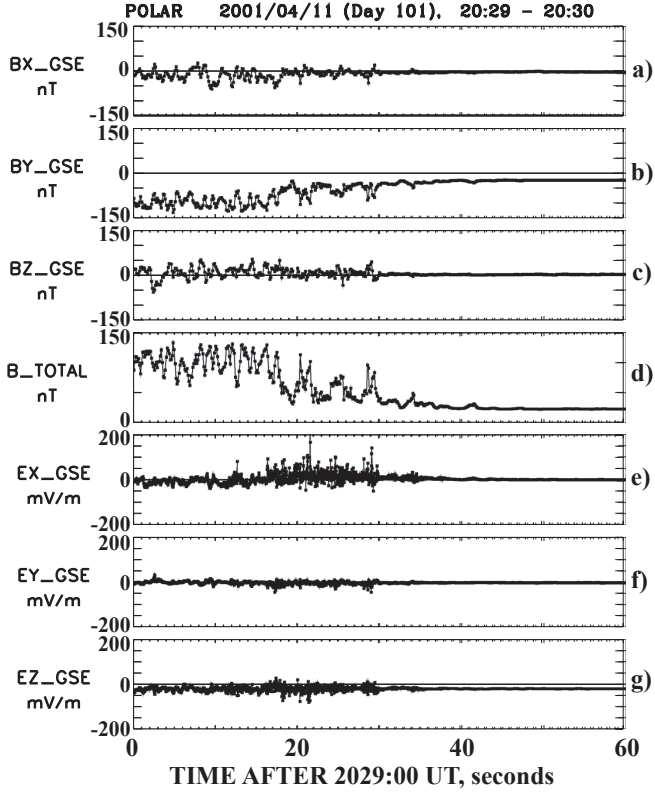}
\caption{Electric and magnetic fields at the April 11, 2001 shock crossing.}\label{fig:currents}
\end{figure}

       Fig. 6 presents 50 msec of data during one of the biggest field events of this crossing.  The potentials of sphere pairs 1-2 and 3-4 were anti-correlated as expected for an external electric field and the time delay between signals in any pair was less than one data point.  This requires that the structure speed was $>$50 km/sec and a speed of 50-100 km/sec will be assumed in the interpretation of the data.  For this range of speeds, the structure that lasted $\sim$15 msec had a thickness of 0.75-1.5 km, which is $\sim$1 electron skin depth and $\sim$10 gyroradii of 10-50 eV electrons.  The perpendicular electric field was as large as 600 mV/m, which is as large as observed anywhere in space.  It pointed generally sunward as would be expected for a field that slows the solar wind ions.  The energy density
       in this field $\epsilon_0 E^2/n k_b T_e$ is approximately 0.12, which implies that the fields are
       highly nonlinear.  The electric potential across this structure was 400-800 volts, depending on the assumed structure speed.  Upstream
shock parameters are given in Table 1.  
\begin{figure}[h]
\centering
 \includegraphics[width=90mm,
	height=110mm, scale=1,clip=true, draft=false]{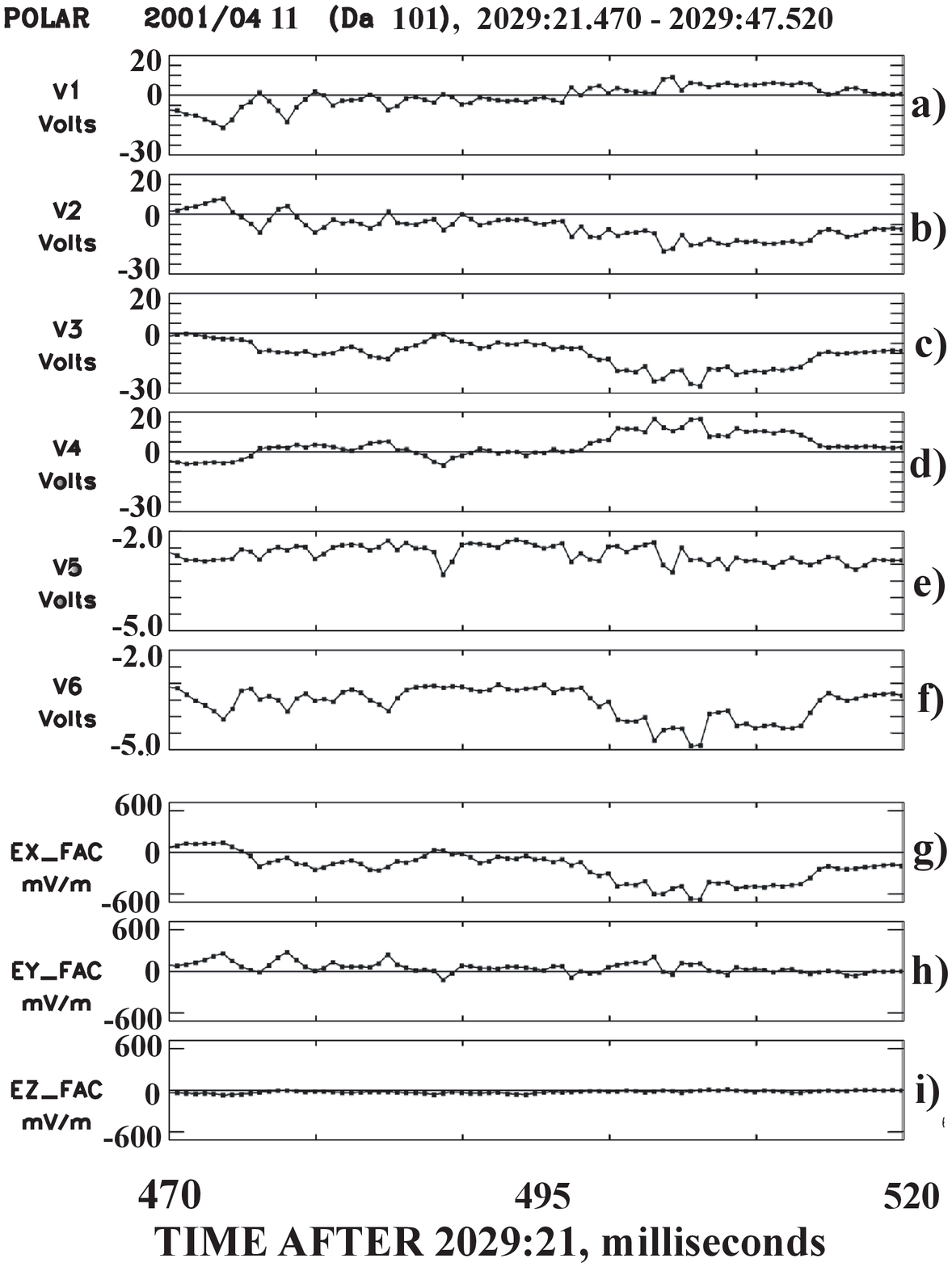}
\caption{Fifty milliseconds of data at the bow shock ramp on April 11, 2001 in similar format to Figure 3.}\label{fig:currents}
\end{figure}
 
The parallel electric potentials of tens of volts and the perpendicular potentials of hundreds to a thousand volts evidently contribute to the total cross-shock
electric potential responsible for slowing the bulk flow and heating the electrons.  The total cross-field potential arises from the accumulation of many short,
spiky electric field structures.

Since these shocks have non-zero parallel electric fields with scale sizes that are comparable to or less than the electron skin 
depth, the electrons are no longer magnetized (i.e. $\vec{E} + \vec{v}_e \times \vec{B} \neq 0$).  This condition is similar to an 
'electron diffusion region' in magnetic reconnection and allows for rapid relative motion of the magnetic field, if the boundary 
conditions provide strong shear.  Furthermore, demagnetized electrons do not obey the first adiabatic invariant and hence some 
part of the electron 'heating' may be due to stochastic motion (rather than purely coherent effects), a point which has been debated
by theorists in the literature \cite{scudder95}.

\begin{table}[h]
  \centering 
  \caption{Shock parameters:  the Alfv\'en Mach number $M_A$, shock tangent angle $\Theta_{bn}$,
  ion and electron plasma beta, and the maximum measured electric fields. }
   \begin{tabular}{|l|l|l|l|l|c|c|}
\hline
  date & $M_A$ & $\Theta_{bn}$ & $\beta_p$ & $\beta_e$ & $|E_\perp|_{max}$  & $E_{\parallel, max}$ \\
\hline
  2001-04-08 & 9.2 & 88$^\circ$  & 0.55 & 0.53 & $\sim$150 {\small mV/m} & $\sim$100 mV/m\\
  2001-04-11 & 8.1 & 85$^\circ$  &0.48 & 0.58 & $\sim$600 mV/m & $\sim$70 mV/m\\
\hline
\end{tabular}
\end{table}

\begin{acknowledgments}
      The authors thank Professor C.T. Russell for providing magnetometer data.  Work at UC Berkeley was carried out 
      under NASA grant NNG05GC72G.
\end{acknowledgments}

\bibliography{shock}
\bibliographystyle{apsrev}
\nocite{*}

\end{document}